\documentclass[conference]{IEEEtran}
\IEEEoverridecommandlockouts{}
\usepackage{amsmath,amssymb,amsfonts}
\usepackage{algorithmic}
\usepackage{graphicx}
\usepackage{textcomp}
\usepackage[dvipsnames]{xcolor}
\usepackage{booktabs}
\usepackage{array}
\usepackage{flushend}

\usepackage{url}

\usepackage[export]{adjustbox}

\ifCLASSOPTIONcompsoc
    \usepackage[caption=false,font=normalsize,labelfont=sf,textfont=sf]{subfig}
\else
    \usepackage[caption=false,font=footnotesize]{subfig}
\fi

\usepackage[separate-uncertainty=true,binary-units]{siunitx}
\usepackage{placeins}
\usepackage{microtype}

\usepackage{hyperref}

\usepackage{cleveref}

\usepackage[noadjust,compress]{cite}

\newcommand{\restore}{ReStore}
\newcommand{\gpicp}{GPI\_CP}
\newcommand{\submit}{\texttt{submit}}
\newcommand{\loadone}{\texttt{load \SI{1}{\percent} data}}
\newcommand{\loadall}{\texttt{load all data}}
\newcommand{\etal}{et al.}
\newcommand{\bigoh}{\mathcal{O}}

\usepackage{pifont}

\newcommand{\cmark}{\ding{51}}%
\newcommand{\xmark}{\ding{55}}%
\newcommand{\greencmark}{{\color{ForestGreen}\cmark}}
\newcommand{\redxmark}{{\color{BrickRed}\xmark}}
\newcommand{\orangecmark}{{\color{orange}(\cmark)}}

\def\BibTeX{{\rm B\kern-.05em{\sc i\kern-.025em b}\kern-.08em
    T\kern-.1667em\lower.7ex\hbox{E}\kern-.125emX}}
    
\usepackage{mathtools}

\DeclarePairedDelimiter\floor{\lfloor}{\rfloor}

\usepackage{tikz}
\newcommand\copyrighttext{%
  \footnotesize \textcopyright 2022 IEEE. Published version: \href{https://doi.org/10.1109/FTXS56515.2022.00008}{10.1109/FTXS56515.2022.00008}~\cite{RestoreConferenceVersion}} 
\newcommand\copyrightnotice{%
\begin{tikzpicture}[remember picture,overlay]
\node[anchor=south,yshift=10pt] at (current page.south) {\fbox{\parbox{\dimexpr\textwidth-\fboxsep-\fboxrule\relax}{\copyrighttext}}};
\end{tikzpicture}%
}

\begin{document}

\title{ReStore: In-Memory REplicated STORagE for Rapid Recovery in Fault-Tolerant Algorithms}

\author{
    \IEEEauthorblockN{
        Lukas Hübner\IEEEauthorrefmark{1}\IEEEauthorrefmark{2}\IEEEauthorrefmark{3},
        Demian Hespe\IEEEauthorrefmark{1}\IEEEauthorrefmark{2},
        Peter Sanders\IEEEauthorrefmark{2},
        Alexandros Stamatakis\IEEEauthorrefmark{2}\IEEEauthorrefmark{3}
    }
    \IEEEauthorblockA{
        \IEEEauthorrefmark{1}
        \textit{primary authors}
    }
    \IEEEauthorblockA{
        \IEEEauthorrefmark{2}
        Karlsruhe Institute of Technology, Karlsruhe, Germany \\
        Email: \{hespe,huebner,sanders,stamatakis\}@kit.edu
    }
    \IEEEauthorblockA{
        \IEEEauthorrefmark{3}
        Heidelberg Institute of Theoretical Studies, Heidelberg, Germany
    }
}


\maketitle
\copyrightnotice

\begin{abstract}
    Fault-tolerant distributed applications require mechanisms to recover data lost via a process failure.
    On modern cluster systems it is typically impractical to request replacement resources after such a failure.
    Therefore, applications have to continue working with the remaining resources.
    This requires redistributing the workload and that the non-failed processes reload data.
    We present an algorithmic framework and its C++ library implementation \restore{} for MPI programs that enables recovery of data after process failures. 
    By storing all required data in memory via an appropriate data distribution and replication, recovery is substantially faster than with standard checkpointing schemes that rely on a parallel file system.
    As the application developer can specify which data to load, we also support shrinking recovery instead of recovery using spare compute nodes.
    We evaluate \restore{} in both controlled, isolated environments and real applications.
    Our experiments show loading times of lost input data in the range of milliseconds on up to 24\,576 processors and a substantial speedup of the recovery time for the fault-tolerant version of a widely used bioinformatics application.
\end{abstract}


\section{Introduction}
With the increasing number of processors in high performance computing clusters, the probability that some processors fail during a computation rises.
Handling such failures constitutes a major challenge for future exascale systems~\cite{Shalf2010}.
For example ORNL's Jaguar Titan Cray XK7 system had on average 2.33 failures/day between August 2008 and 2010~\cite{Gamell2014}.
In upcoming systems, we expect a hardware failure to occur every 30 to 60 minutes~\cite{Cappello2014,Dongarra2015,Snir2014}.
To recover from such a failure, an important step is to restore the lost data that the failed processors were working on.
To save the current state of a program's data, applications write \emph{checkpoints} which can be reloaded after a process failure.
Checkpointing libraries usually write their checkpoints to a parallel file system (PFS)~\cite{Agarwal2004,BautistaGomez2011,Shahzad2019,Nicolae2019}, implying slow recovery due to low disk access speeds and because many processors simultaneously access the same resources.
Many checkpointing libraries also assume the nature of the failures to leave the machine in a state where the process can simply be started again, or they assume that enough spare resources are kept idle to start a new process for replacing the failed one~\cite{Agarwal2004,BautistaGomez2011,Shahzad2019,Nicolae2019,Teranishi2014,Moody2010,Besta2014,Gamell2014,Lu2005,Bartsch2017}.
Under this assumption, a re-spawned process can simply read exactly the data of the failed process.
In the case of the new process being located on the same compute node as the failed one, the checkpoint can even be read from a local disk.
To the best of our knowledge there exists no general purpose checkpointing solution that allows for in-memory recovery without requiring spare resources.

\subsubsection*{Contribution and Structure.}
We introduce \restore{}, an in-memory checkpointing library that is optimized for recovery speed (in contrast to checkpoint creation speed).
This is especially important for data which the program never or rarely changes but has to be redistributed after every failure.
We do not assume that spare resources are available.
Instead, \restore{} enables recovery in an application that continues its execution only with the processes that are still alive.
While this approach requires a more involved recovery mechanism and strategic data distribution, it saves resources because all available processors can participate in the application's useful computations from the beginning.
Keeping the checkpoints in-memory avoids the bottlenecks involved in a PFS and allows high scalability.
The remainder of this paper is structured as follows: We introduce the concepts used in this paper in \Cref{sec:prelims} and provide an overview of existing checkpointing libraries and other related work in \Cref{sec:related-work}.
We explain our general framework and data distribution in \Cref{sec:main-section}.
In \Cref{sec:implementation} and~\ref{sec:experiments} we present the implementation of our open source C++ library and the experimental results, respectively.
We conclude in \Cref{sec:conclusion} and outline future work.

\section{Preliminaries}\label{sec:prelims}

\begin{table*}[!t]
    \renewcommand{\arraystretch}{1.3}
    \caption[]{
        Comparison of checkpointing libraries. 
        \rm See \Cref{sec:related-work} for details.
        \footnotemark[1]The program needs to allocate spare nodes, which participate in the computation only in case of a failure.
        \footnotemark[2]The program needs to allocate spare nodes and nodes used purely for checkpointing.
        \footnotemark[3]The maintenance state unclear (\Cref{sec:competitor-reproducibility}).}
    \label{tbl:related-work}
    \centering
    \begin{tabular}{rcccccc} 
        & \textbf{ftRMA}~\cite{Besta2014} & \textbf{Fenix}~\cite{Gamell2014} & \textbf{SCR}~\cite{Moody2010} & \textbf{Lu}~\cite{Lu2005} & \textbf{\gpicp{}}~\cite{Bartsch2017} & \textit{\textbf{ReStore}} \\
        \toprule{}
        \textbf{Features} \\
        in-memory checkpointing  & \greencmark{} & \greencmark{} & \redxmark{} & \greencmark{} & \greencmark{}& \greencmark{}             \\
        substituting recovery & \greencmark{} & \greencmark{} & \greencmark{} & \greencmark{} & \greencmark{} & \greencmark{}             \\
        shrinking recovery & \redxmark{} & \redxmark{} & \redxmark{} & \redxmark{} & \redxmark{} & \greencmark{}             \\
        all nodes participate in computation  &  \redxmark{}\footnotemark[2] & \orangecmark{}\footnotemark[1] & \orangecmark{}\footnotemark[1] & \redxmark{}\footnotemark[2] & \orangecmark{}\footnotemark[1] & \greencmark{} \\
        programming model        &               MPI RDMA & MPI & MPI & MPI & PGAS/GPI & MPI \\
        \textbf{Reproducibility} \\
        source-code available & \greencmark{} & \greencmark{}  & \greencmark{} & \redxmark{} & \greencmark{} & \greencmark{} \\
        still maintained (2022)  & \redxmark{} & {\color{orange} ?}\footnotemark[3] & \greencmark{} & \redxmark{} & \redxmark{} & \greencmark{}\\
        other reproducibility issues & Cray-only &  author-provided         & & & requires libiverbs, GPI-2 & \\
                                     &            &  examples segfault       & & & and passwordless ssh-login & \\
                                     &            &                          & & & on all compute nodes      & \\
    \end{tabular}
\end{table*}
In distributed memory parallel programs using the \emph{Message Passing Interface} (MPI), \(p\) processes (or \emph{processing elements} (PEs)) run on multiple machines (or \emph{nodes}) and communicate via messages sent over the network.
We consider two important factors for evaluating the running time of such parallel algorithms: The \emph{bottleneck} number of messages sent and received, and the \emph{bottleneck communication volume}.
The bottleneck number of messages sent and received describes the maximum number of messages sent or received by a single PE\@.
This influences performance, as there is a startup overhead (latency) for establishing a connection associated with each message.
As a result, sending or receiving data from one PE to another in a single message is usually faster than splitting that data into many parts and sending each part to a different receiver.
The bottleneck communication volume describes the maximum amount of data sent or received by a single PE and represents a point on the critical path of the application.

As faults, we consider the case that one or multiple PEs suddenly stop working and do not contribute to the computation anymore (which we will refer to as \emph{failed}).
The upcoming MPI 4 standard will include fault-tolerance mechanisms and an implementation called ``ULFM'' is available for OpenMPI~\cite{Bland2013}.
Following a fault, the application has to redistribute the work formerly performed by a failed PE using either the \emph{shrink} or the \emph{substitute} strategy~\cite{Ashraf2018}.
Under the \emph{substitute} strategy, a replacement PE takes over the work previously performed by the failed PE\@.
This circumvents the need for re-balancing the workload and simplifies loading the required data.
However, reserving idle processors for this purpose constitutes a waste of resources.
In the \emph{shrink} strategy, the program's load balancer (re)distributes the work performed by the failed PE among the remaining (or \emph{surviving}) PEs. 
This strategy does therefore not require spare PEs but requires reloading fractions of the data on many or even all PEs.
While the number of failures an algorithm can tolerate using the \emph{substitute} strategy is limited to the number of spare PEs; this limitation does not apply to the \emph{shrink} strategy~\cite{Ashraf2018}.

\section{Related Work}\label{sec:related-work}

\begin{figure*}[!t]
    \centering
    \includegraphics{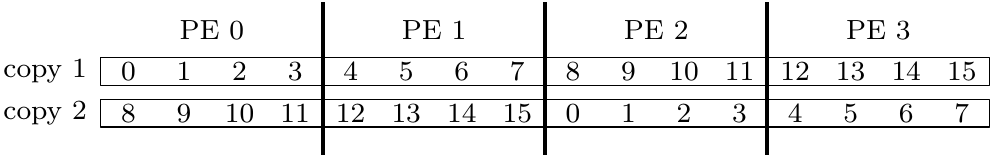}
    \caption{Example showing the data distribution of the copies stored with
        \restore{} for \(p = 4\) PEs, \(n = 16\) data blocks, and \(r = 2\)
        copies.\label{fig:general-framework}}
\end{figure*}

Scientific applications are increasingly implemented to tolerate faults.
Examples include a numeric linear equation and partial equation solver~\cite{Ali2016}, a plasma simulation~\cite{Obersteiner2017}, a molecular dynamics simulations~\cite{Laguna2016}, a Fast Fourier Transformation~\cite{Engelmann2003}, and an algorithm for phylogenetic inference~\cite{Huebner2021}.
The three main techniques for implementing fault-tolerant algorithms are Algorithm-Based Fault-Tolerance~\cite{Vijay1997,Bosilca2008}, restarting failed sub-jobs~\cite{Memishi2016}, and checkpointing/restart~\cite{Kohl2019,Huebner2021}.
Checkpointing/restart can be further subdivided into coordinated and uncoordinated checkpointing.
In coordinated checkpointing, the program synchronizes before creating the checkpoint in a distributed-manner.
This ensures that there are no messages in-flight and the program's state is therefore well-defined.
Gavaskar and Subbarao recommend coordinated checkpointing for the high-bandwidth, low-latency interconnections of modern HPC systems~\cite{Gavaskar2013}.
Checkpointing libraries can save their checkpoint either to a (possibly network attached) disk or to the compute node's main memory (``diskless'')~\cite{Plank1998}.
Checkpointing libraries which save their checkpoints to disk include, for example, the algorithm presented by Agarwal \etal~\cite{Agarwal2004}, FTI~\cite{BautistaGomez2011}, CRAFT~\cite{Shahzad2019}, SCR~\cite{Moody2010}, and VeloC~\cite{Nicolae2019}.
As the number of nodes per parallel program execution continues to grow, the congestion on the PFS increases -- resulting in a bottleneck and reduced checkpointing performance~\cite{Gossman2021,Herault2019}.
Examples for in-memory checkpointing libraries include ftRMA~\cite{Besta2014}, Fenix~\cite{Gamell2014}, \gpicp{}~\cite{Bartsch2017}, and the algorithm described by Lu~\cite{Lu2005} (\Cref{tbl:related-work}).
All of these employ the substitute strategy and therefore rely on the availability of replacement nodes, if we want to continue the computation in case of node failure.
This implies that some nodes are allocated to the job but not available for computation.
Some algorithms additionally designate some compute nodes as pure checkpointing nodes, which are neither participating in the computation nor available as spares.
Ashraf \etal~\cite{Ashraf2018} describe an implementation of a fault-tolerance mechanism for a specific application which is able to checkpoint to memory and recover in a shrinking setting.
This is, however, not a general-purpose checkpointing library but application-specific.
Erasure codes are often used to reduce the file-size or memory footprint of checkpoints~\cite{Lu2005,Besta2014,BautistaGomez2011}.

Other areas where replication approaches similar to the one presented in this paper are used are distributed fault-tolerant file systems like early versions of the Hadoop Distributed File System~\cite{shvachko2010hadoop} or distributed processing frameworks like Apache Spark~\cite{zaharia2012resilient}. However, these target very different use cases and sometimes only support very basic replication like storing each PEs data on a single partner PE.

\subsection{Reproducibility Study}
\label{sec:competitor-reproducibility}

In the following, we describe our attempts to replicate the results of competing tools.
We provide a visual summary of these in \Cref{tbl:related-work}.

The ftRMA~\cite{Besta2014} tool has not been maintained since 2014 and relies on the Cray-only foMPI library which has also not been further maintained since 2014.
The authors confirmed (pers.~comm.~30\@.~June~2022) that the current code exclusively works on Cray systems and is no longer being actively maintained.
Although the authors suggested that ftRMA could -- in principle -- be ported to a non-Cray system, taking into account the unmaintained code base comprising 513 calls to foMPI functions, this would incur a prohibitive programming effort with uncertain outcomes.
Further, as ULFM currently provides ``little support for fault tolerance'' with respect to RMA calls~\cite{Bouteiller2019}, deploying ftRMA would be bound to fail using a current fault-tolerant MPI implementation.

With respect to the Fenix tool, there has only been a  single commit to its repository within the past six months.
In addition, the author's automated testing on GitHub failed for this commit.
We thus denote Fenix's maintenance status as being ``unclear'' \orangecmark{} in \Cref{tbl:related-work}.
The author-provided Fenix examples~\cite{Gamell2014} fail with a segmentation fault.
As Fenix does currently not support restoring the data that was checkpointed on a different rank, setting up an experimental comparison is challenging.
The authors did not respond to our e-mail requesting assistance.

SCR~\cite{Moody2010} has \(> 50\) commits on 20 distinct days during the past six months.
Hence, we consider that it is still being maintained. 
SCR supports \emph{caching} checkpoints on a RAM-disk.
These checkpoints, however, have to be transferred to the parallel file system such as to become available upon rank failure.
We therefore do not consider SCR to be an in-memory checkpointing library in the context of node failures.

The source code of Lu~\cite{Lu2005} is not available, and the author can not be contacted, as they did not provide a contact e-mail address on their publications.

The git repository of \gpicp{}~\cite{Bartsch2017} has only a single commit from six years ago; we thus also consider that it is no longer being maintained.
As we do not have access to an HPC system where GPI-2 (a library for the Partitioned Global Address Space (PGAS) programming model) is supported, and its dependency \texttt{libiverbs} has to be installed by a system administrator, we are unable to compare \gpicp{} against ReStore.

\section{In-Memory Replica for Fast Recovery}\label{sec:main-section}

\restore{} allows application developers to store redundant copies of their data in-memory.
In case of a failure, the surviving PEs can invoke a recovery routine to load all or parts of the data lost during the failure.

The remainder of this section is structured as follows: In \Cref{sec:general} we introduce our general framework for maintaining redundant copies of the user-supplied data in memory as well as the algorithm used for recovery.
\Cref{sec:permutations} expands on the distribution of copies by adding random permutations that accelerate the recovery algorithm.
We analyze the memory usage of our proposed data distribution in \Cref{sec:memory-usage} and the probability of irrecoverable data loss in \Cref{sec:probability-of-idl}.
In \Cref{sec:restore_redundancy} we describe a currently unimplemented approach to restore the level of redundancy after a failure.

\subsection{General Framework}\label{sec:general}

The main idea of \restore{} is to store \(r\) copies of the data on different nodes.
By storing them such that it is unlikely for all copies of one data element to fail at once, there will most likely (\Cref{sec:probability-of-idl} and \Cref{sec:data-loss-simulation}) be copies left to recover from.
The application programmer can store data into ReStore using the \emph{submit} and retrieve data form ReStore using the \emph{load} function.
To make the data addressable, we divide it into blocks where each block has a unique identifier.

Let \(n\) be the number of data blocks.
In its most basic form (\Cref{fig:general-framework}), for \(k \in \left[0, r\right)\) we store the block with ID \(x\) on PEs \( L(x, k) = \floor*{\frac{xp}{n}} + k \cdot \frac{p}{r} \mod p \).
Under this distribution, we expect the copies of a block to not be stored on the same physical node/case/rack in most cluster setups.
This decreases the probability of loosing all copies of a block, as failures of PEs in the same node/case/rack are more likely to occur than a simultaneous failure of unrelated PEs~\cite{BautistaGomez2011}.
In Section~\ref{sec:permutations} we explore a change to this basic distribution scheme that allows for faster recovery.

\begin{figure*}[!t]
    \centering
    \includegraphics{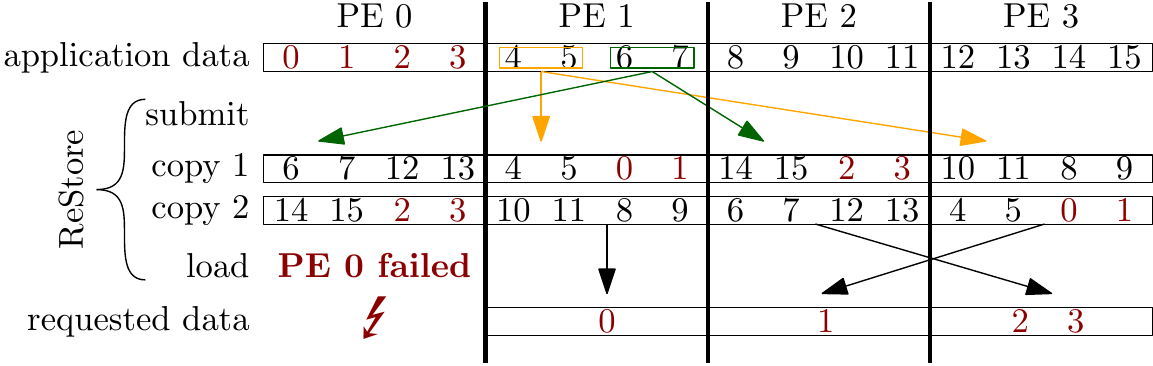}
    \caption{Example showing the submit and load operations as well as the data distribution of the copies with a random permutation for \(p = 4\) PEs, \(n = 16\) data blocks, \(r = 2\) copies, and \(s_\mathit{pr} = 2\) blocks per permutation range.
        The first row shows the data submitted by the application.
        As an example, the orange and green arrows show the data \restore{} sends from PE \(1\) to the target PEs which hold copies of the received data.
        After PE \(0\) fails, the application requests the data shown in the last row
        (dark red in all occurrences) which is served by \restore{} as shown with the black arrows.\label{fig:block-permutations}}
\end{figure*}

During recovery, when PE \(i\) requests to load block \(j\), we choose one of the surviving PEs that hold block \(j\) at random to serve the request.
If the user requests multiple successive blocks which are stored on the same set of PEs, we choose one PE to serve all requests.
This strategy minimizes the bottleneck number of messages received.
Next, we distribute the requested data using a custom sparse all-to-all communication.

\subsection{Breaking Up Access Patterns for Faster Recovery}\label{sec:permutations}

The goal of distributing data copies as described in \Cref{sec:general} is primarily to preserve the ability to recover from a fault. In the following, we explore how to adapt the data distribution such that it accelerates data recovery while preserving the level of failure resilience.

Assume a failed PE \(i\) which worked on data blocks \(\left[i \frac{n}{p}, (i+1) \frac{n}{p}\right)\), where \(n\) is the total number of data blocks submitted to \restore{}. 
If the application redistributes the lost data evenly to all surviving PEs, we would ideally want a dedicated sending PE for each receiver, resulting in a bottleneck communication volume of \(\frac{n}{p^2}\) and a bottleneck number of messages received of \(1\).
With the data distribution from \Cref{sec:general}, only the surviving subset of the \(r \ll p\) PEs that hold copies of these blocks act as sources, resulting in a bottleneck communication volume of \(\frac{n}{pr}\).
We can alleviate this issue by evenly distributing the copies of data blocks \(\left[i \frac{n}{p}, (i+1) \frac{n}{p}\right)\) among multiple PEs by randomly permuting the block identifiers. 
For a random permutation \(\pi\) and \(k \in \left[0, r\right)\), the block \(x\) is stored on PEs
\[
    L(x, k) = \floor*{\frac{\pi(x) \cdot p}{n}} + k \cdot \frac{p}{r} \mod p
\]
If the user requests blocks \(\left[i \frac{n}{p}, (i+1) \frac{n}{p}\right)\), more PEs have parts of the data and can send them to the requesting PE\@. 
This approach, however, can lead to a large bottleneck number of messages being sent and received: If a PE requests \(\frac{n}{p^2}\) data blocks, these blocks can reside on up to \(\min(\frac{n}{p^2}, p)\) different PEs.
To mitigate this, we group the data into \emph{permutation ranges} of size \(s_\textit{pr}\).
We then apply a random permutation to these permutation ranges instead of
individual data blocks (\Cref{fig:block-permutations}).
If by a fault on PE \(i\), data blocks \(\left[i\frac{n}{p}, (i+1) \frac{n}{p}\right)\) need to be redistributed, these correspond to permutation ranges \([\frac{i \cdot n/p}{s_\textit{pr}}, \frac{(i+1) \cdot n/p}{s_\mathit{pr}})\). 
If we request the data to be evenly distributed among the \(p - 1\) surviving PEs such that PE \(j\) receives blocks \([i\frac{n}{p} + j \frac{n}{p \cdot (p-1)}, i\frac{n}{p} + (j+1) \frac{n}{p \cdot (p-1)})\), only \(\frac{n/(p \cdot (p-1))}{s_\mathit{pr}}\) PEs send to every receiving PE. 

The best choice of \(s_\mathit{pr}\) and whether to use permutations at all depends on the data distribution, the expected amount of data lost by a fault, and the application's recovery strategy, but also on the frequency of checkpoint creation as submitting data with permutations enabled results in a more dense communication pattern. \Cref{sec:data-loss-simulation} shows how we experimentally chose a good value for~\(s_\mathit{pr}\).

Note that with this data distribution, we always have sets of \(\frac{n}{s_\mathit{pr}p}\) permutation ranges that are stored together for all \(r\) copies (e.g., blocks \(6,5,12,13\) in \Cref{fig:block-permutations} are always stored together -- once on PE~\(0\) and once on PE~\(2\)).
This means that for all copies of any permutation range whose first copy is stored on PE~\(i\) to become lost, exactly the set of \(r\) PEs~\(i + k \cdot \frac{p}{r}, k \in \left[0, r\right)\) has to fail.
One could also opt for a different approach -- for example, using a distinct permutation for each copy.
In this case, no sets of permutation ranges will always be stored together.
So in order for any permutation range whose first copy resides on PE~\(i\) to be lost, it is sufficient if \emph{any} of the \(\frac{n}{s_\mathit{pr}p}\) sets of PEs of size \(r\) fail that hold the copies of one of the permutation ranges.
In \Cref{sec:probability-of-idl} we analyze the probability of irrecoverable data loss under our proposed data distribution.

\subsection{Memory Usage}
\label{sec:memory-usage}

Other fault-tolerance libraries~\cite{Lu2005,Besta2014,BautistaGomez2011} often use erasure coding -- for example the Reed-Solomon code~\cite{Reed1960} -- to reduce their memory footprint.
This works for example by not storing the replicas \(A'\) and \(B'\) of two blocks \(A\) and \(B\) but rather the XOR of these blocks \(A \oplus B\).
We decide against using erasure coding as a means to reduce the memory footprint of our checkpoints, as this would incur additional messages upon checkpoint creation and recovery as well as a substantial computational overhead~\cite{Chiueh1996}.
We therefore trade reduced communication overhead for increased memory consumption. 

As in \Cref{sec:general}, let \(n\) be the number of data blocks, \(r\) be the number of replicas and \(p\) be the number of processes.
On each PE \restore{} requires main-memory to store \(\frac{rn}{p}\) data blocks for the replicated storage.
The memory requirement is doubled during submission as we require additional space for the send and receive buffers.
During recovery, an additional copy of all data being sent and received is stored on each PE\@.
We verified these formulas empirically (data not shown).
A plethora of applications exist for which the amount of memory for the input data {\em and} the data that need to be checkpointed fit in memory \(r\) times. Examples include RAxML-NG~\cite{Kozlov2019, Huebner2021}, $k$-means, and page-rank.\footnote{We implemented fault-tolerant version for all three of these using \restore{} and show running times for RAxML-NG and k-means in \Cref{sec:applications}.}
For example, RAxML-NG is memory bandwidth bound~\cite{Kozlov2018}.
Hence, using additional cores with their associated larger cache memory capacity can even yield super-linear speedups due to increased cache-efficiency.
Such applications can therefore substantially benefit from the reduced communication and computational overhead to create and restore a checkpoint without erasure codes.

\subsection{Probability of Irrecoverable Data Loss.}
\label{sec:probability-of-idl}

Let \(r\) be the replication level and \(p\) be the number of PEs.
In this analysis, we assume that \(r|p\) (\(r\) divides \(p\)).
This constitutes a reasonable assumption for current two socket systems that exhibit an even number of cores per socket and \(r=4\).
If \(r|p\), the PEs are divided into \(g = \frac{p}{r} \) groups, with all PEs in a respective group storing the same data.
Thus, if and only if all \(r\) PEs in a specific group fail, we will not be able to recover a part of the data.
We denote such an event as Irrecoverable Data Loss (IDL).
Let \(f\) be the number of failed PEs.
There is exactly one possibility to draw \(r\) out of \(r\) PEs belonging to a single group.
The number of possibilities to draw the remaining \(f - r \) failed PEs among the remaining PEs such that they do \emph{not} belong to the given group is \(\binom{p - r}{f - r}\).
The overall number of possibilities to draw \(f\) PEs from the \(p\) PEs that are still alive at program start is \(\binom{p}{f}\).
The probability that, given \(f\) failures, all processes of a \textit{given} group fail is thus \({1 \cdot \binom{p - r}{f - r}} / {\binom{p}{f}}\).
When generalizing this equation to the probability of all processors of \emph{at least one} group failing, we have to apply the inclusion-exclusion principle to avoid counting the same combination multiple times.
We thus obtain the following equation for the probability of an IDL at failure \(f\) or any failure before:
\[
    P_\mathrm{IDL}^\le(f) = \sum^g_{j=1} {(-1)}^{j+1} \binom{g}{j} \frac{\binom{p - j r}{f - j r}}{\binom{p}{f}}
\]
The probability of an IDL at exactly failure \(f\) is thus:
\[
    P_\mathrm{IDL}^=(f) = P_\mathrm{IDL}^\le(f) - P_\mathrm{IDL}^\le(f - 1)
\]
The expected number of failures until an IDL occurs is:
\[
    E[\textrm{Failures until IDL}] = \sum_{f = r}^{p} P_\mathrm{IDL}^=(f) \cdot f
\]

For small $f$, the \emph{approximate} probability of all PEs of any group failing is given by $P_\text{IDL}^\text{approx.}(f) = g \cdot (f/p)^r$.
Solving for the fraction of PEs that fail $f/p$ such that $P_\text{ID}^\text{approx.} = 1$ yields $f/p = (r/p)^{(1/r)} \in \mathcal{O}(p^{-1/r})$ for a fixed $r$.

In \Cref{sec:data-loss-simulation} we simulate node failures using the actual data distribution to verify these formulas.

\subsection{Recovering Lost Replicas After a Node Failure}
\label{sec:restore_redundancy}


To further increase the resilience of our framework, we introduce an approach to restore replicas that were lost upon a failure while keeping all other replicas in place.
That is, we do not need to redistribute any replicas that reside on surviving nodes.
As in the previous sections, let \(n\) be the number of data blocks, \(r\) be the number of replicas per block, and \(p\) be the number of nodes.

We draw a different random permutation $\rho_x$ of $[0, p-1]$ (or long, non-repeating random sequences of nodes) for each block $x$ and place the replicas of $x$ on the first $r$ alive nodes of that permutation. When a node dies, we copy all replicas that this node held to the next node in each replicas permutation. We can refine this approach to attain a perfectly balanced initial data distribution and reduce the probability of an IDL (\Cref{sec:permutations}): We initially place the first \(r\) replicas (\(L(x, 0), L(x, 1), \ldots, L(x, r-1)\)) deterministically as described in \Cref{sec:general}.
That is, the data distribution is given by 
\[
L(x, k) = \begin{cases}
    \text{as described in \Cref{sec:general}}, & \text{if } k < r \\
    \rho_x(k),            & \text{else}
\end{cases}
\]
See the Appendix for different options to draw such a sequence of nodes.
Following this data distribution, we can compute the ranks on which we store a given block in \(\bigoh(r + f)\) time and \(\bigoh(1)\) space where \(r\) is the number of replicas per block and  \(f\) is the number of node failures. In order to keep recovery fast, we can apply this technique on a permutation-range-level rather than on individual blocks as explained in \Cref{sec:permutations}.

\section{Implementation}
\label{sec:implementation}
\begin{figure*}[!t]
    \subfloat[Simulation (\Cref{sec:data-loss-simulation})\label{fig:dataloss-simulation-all-r}]{
        \centering
        \includegraphics[valign=t]{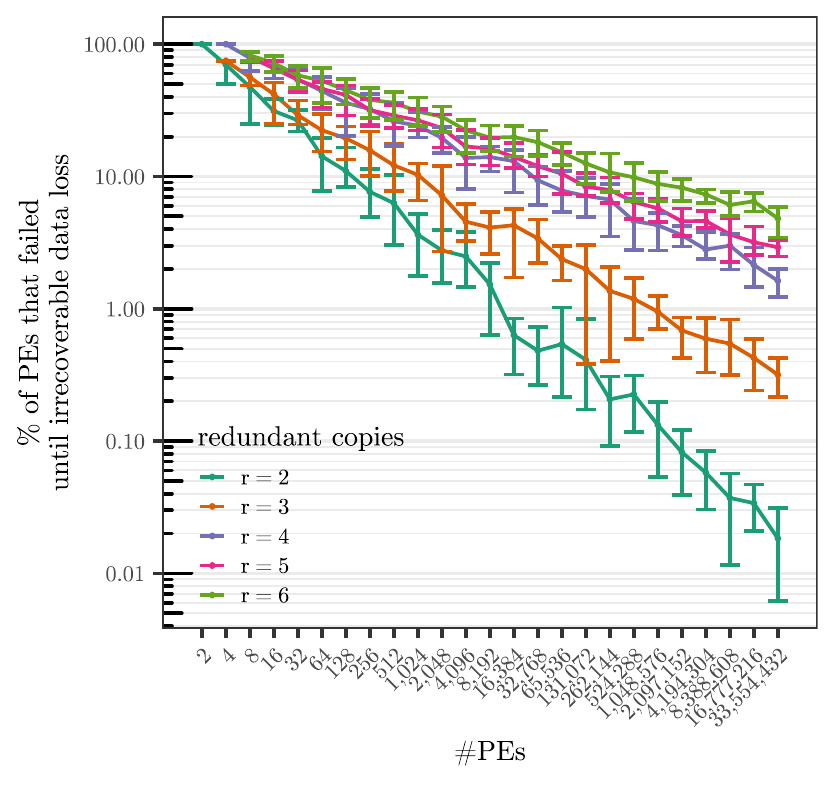}
    }
    \subfloat[Theory (\Cref{sec:probability-of-idl})\label{fig:dataloss-simulation-theory}]{
        \centering
        \includegraphics[valign=t]{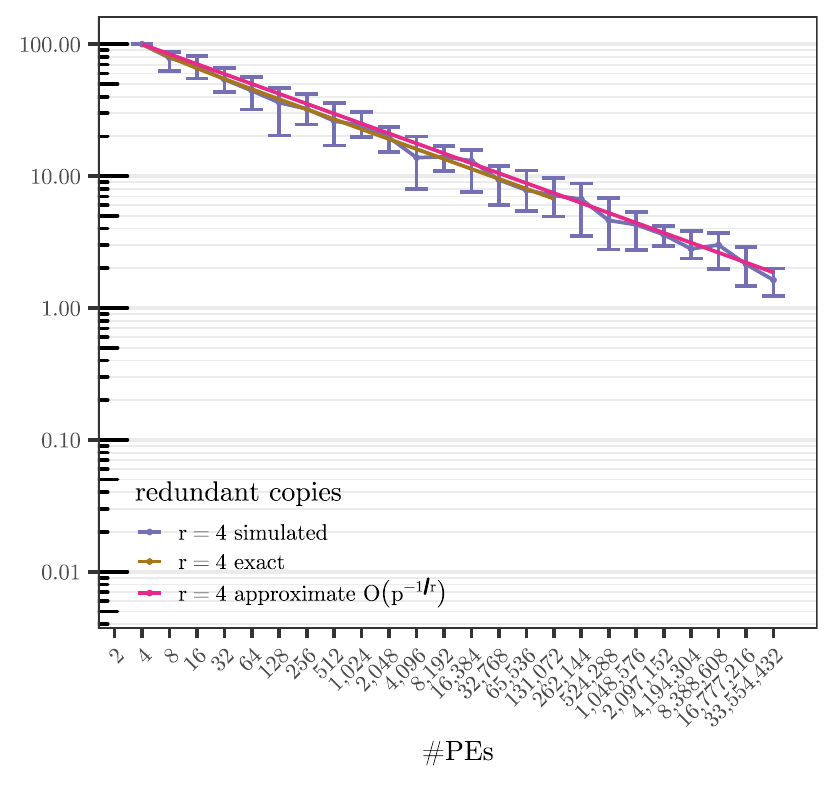}
        \vphantom{images/failure-until-irrecoverable-data-loss}
        \vspace{50mm}
    }
    \caption{
        Percentage of failed PEs until all redundant copies of one data block are lost.
        (a) Simulation of the data distribution described in \Cref{sec:general}.
        We continue simulating the failure of random PEs until there is at least one data block with no remaining copies on the surviving PEs.
        (b) Comparison of the probability given by the equations in \Cref{sec:probability-of-idl} and the simulated values from (a).
        \label{fig:dataloss-simulation}
    }
\end{figure*}

We implement \restore{} as a C++ library\footnote{\url{https://github.com/ReStoreCpp/ReStore}} using the User Level Failure Mitigation (ULFM) proposal implementation~\cite{Bland2013}.
Application programmers submit their data to \restore{} by writing their serialized data blocks to a memory location supplied by the library or using ReStore's interface for already serialized data.
After a failure, they can request data blocks by passing a list of ranges of block identifiers to \restore{}.
This can be done in two ways: Either by providing the full list of requested block IDs on all PEs or by providing exactly those ID ranges each individual PE needs on exactly that PE\@.
By using the first approach, no communication is required to determine which PE serves which request.
When using the second approach, the receiving PE will determine which PE should send each requested data block.
Then, a sparse all-to-all communication is performed to issue the requests to the sending PEs.
Preliminary experiments showed that the latter method performs substantially better because the full list of requests usually scales with the number of PEs in the application, slowing down the first approach.
Thus, for all experiments in Section~\ref{sec:experiments}, we employ the second approach.
As \restore{}'s implementation currently focuses on fast recovery, it provides only an interface for submitting data once and thus is currently not suitable for repeatedly checkpointing changing data. This is sufficient for many fault-tolerant applications -- two of which are demonstrated in \Cref{sec:applications}.

\section{Experimental Evaluation}\label{sec:experiments}

In this section we present the results of our experimental evaluation.
We present the experimental environment in \Cref{sec:environment}.
In \Cref{sec:isolated-evaluation} we evaluate \restore{}'s fault resilience and performance in isolation.
In \Cref{sec:applications} we show how \restore{} performs when used in two fault-tolerant applications: A simple \(k\)-means algorithm and a complex bioinformatics application used by thousands of researchers.
Finally, in \Cref{sec:disks} we compare \restore{} to reading from a parallel file system (PFS) -- which represents a lower bound for checkpointing libraries using the PFS as storage -- as well as the reported running times by other checkpointing libraries.

\subsection{Environment and Experimental Setup}\label{sec:environment}
We run our experiments on the SuperMUC-NG super computer.\footnote{\url{https://doku.lrz.de/display/PUBLIC/SuperMUC-NG}}
Each node consists of two Intel Skylake Xeon Platinum 8174 processors with \(24\) cores and \(96\) GB of memory each, connected via an OmniPath network with a bandwidth of \(100\) Gbit/s.
The operating system is SUSE Linux Enterprise Server 15 SP1 running Linux Kernel version 4.12.14-197.78. 
We compile our benchmark applications using gcc version 10.2.0 with full optimizations enabled (\texttt{-O3}) and all assertions disabled.
Unless otherwise stated, we communicate using OpenMPI version 4.0.4.
We verify that our implementation \emph{does work} if nodes actually fail and communication is recovered with ULFM as part of our fully automated unit tests.
The current version of ULFM, however, is not stable enough to conduct reliable performance benchmark experiments.
For example, processes may be reported incorrectly as failed or recovery may result in two separate groups of nodes that each assume that the other group has failed.
We reported this behavior on the ULFM mailing list and the authors of ULFM reproduced and confirmed the bug.\footnote{George Bosilca. Post \texttt{pbSToy94RhI/xUrFBx\_1DAAJ} on the ULFM mailing list.}
Additionally, most communication and fault tolerance mechanisms are currently slow (see Hübner \etal~\cite{Huebner2021} for details).
We expect these issues in ULFM to be fixed once fault-tolerance is part of the MPI standard and more resources are allocated to implementing these features.
In our performance benchmarks, we thus use OpenMPI and simulate failures by removing processes from the calculation using \texttt{MPI\_Comm\_split} and replacing other required fault recovery steps by functionally similar ones (e.g., replacing \texttt{MPIX\_Comm\_agree} with \texttt{MPI\_Barrier}s).
As \restore{} currently only supports submitting data once, all experiments shown in this section submit only their input data.
This is a restriction in the API, not the underlying algorithm, and will be removed in future work.

All plots show results for \(10\) repetitions per experiment.
Plots depict the mean with error bars for the 10th and 90th percentile.

\subsection{Isolated Evaluation}\label{sec:isolated-evaluation}

In this section we explore \restore{} in isolation.
We first choose the number of redundant copies (\Cref{sec:general}).
Next, we analyze \restore{}'s performance and experimentally optimize the size of permutation ranges (\Cref{sec:permutations}).

\subsubsection{Number of Redundant Copies.}\label{sec:data-loss-simulation}
\Cref{fig:dataloss-simulation-all-r} shows the result of a simulation of our data distribution: We continue simulating the failure of random PEs until at least for one data block no copies remain on the surviving PEs.
We can see that even for \(2^{25}\) PEs, more than \SI{1}{\percent} of all PEs have to fail until we can no longer recover all data when using \(r=4\) redundant copies.
Even in the event of an irrecoverable data loss, the program will not crash, but will have to merely reload the input data from disk.
For applications running on fewer PEs, an even smaller number \(r\) of redundant copies is sufficient to yield data loss unlikely.
For all further experiments we therefore set the number of redundant copies to \(r:=4\).
We compare the values obtained by applying the formula in \Cref{sec:probability-of-idl} with the values obtained by simulation in \Cref{fig:dataloss-simulation-theory} showing that our theoretical formula matches the simulation very closely.

\subsubsection{Performance.}\label{sec:parameter-optimization}

\begin{figure*}[!t]
    \subfloat[Optimizing the number of bytes per permutation range]{ 
        \centering
        \includegraphics[valign=t]{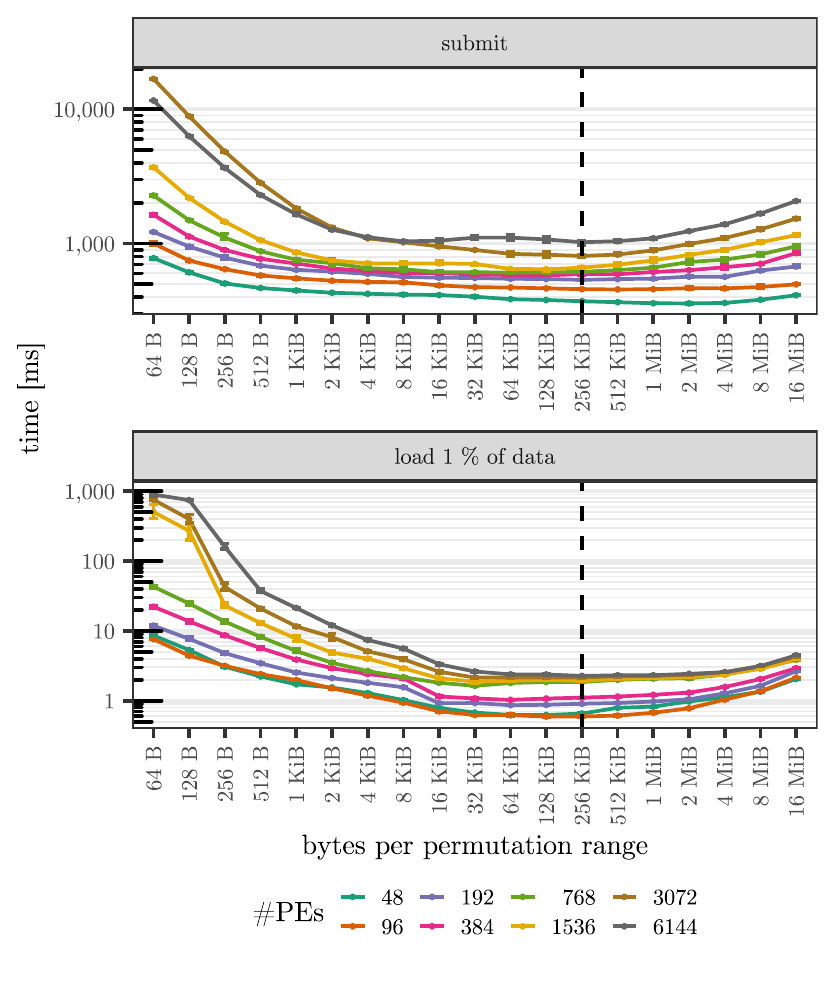}
        \label{fig:permutation-range-size}
    }
    \subfloat[Randomized vs consecutive IDs]{
        \centering
        \includegraphics[valign=t]{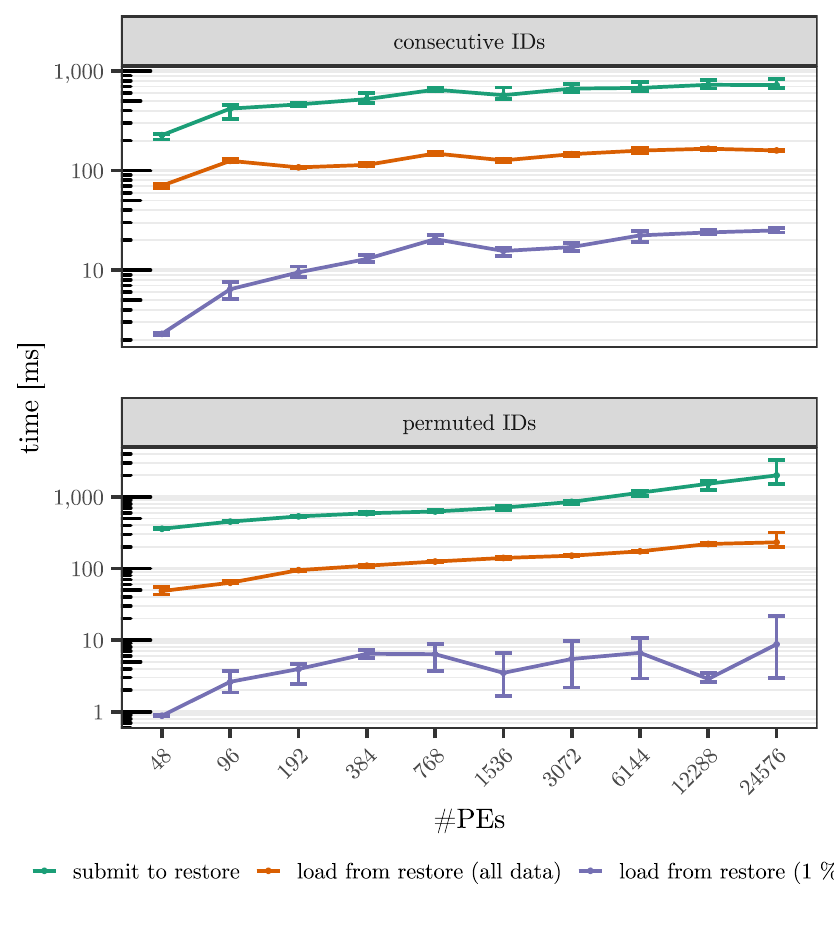}
        \vphantom{images/determine-optimal-bytes-per-permutation-range}
        \label{fig:id-randomization-on-or-off}
    }
    \caption{ (a) Influence of the number of bytes per permutation range on the running time of submitting to and loading from ReStore.
    (b) Weak scaling experiment (\SI{16}{\mebi\byte} per PE) of our three benchmark operations with and without ID randomization. We copy all data over the network -- i.e., no rank holds a copy of its requested data in its local part of the ReStore storage.} 
\end{figure*}

For all experiments in this section we use data blocks of size \SI{64}{\byte} and
\SI{16}{\mebi\byte} of data per PE\@.
We show results for three different operations: In the \submit{} operation we pass \SI{16}{\mebi\byte} on all PEs to \restore{}'s submit function.
In \loadone{} we load the data submitted by \SI{1}{\percent} of the PEs with contiguous data block IDs evenly across all PEs.
So for \(n\) total data blocks, we pick a random starting PE \(i\) and request data blocks \(i \cdot n/p\) to \((i + 0.01 \cdot p) \cdot n/p\).
This simulates the requests expected if \SI{1}{\percent} of PEs fail at once.
In \loadall{} we load all data stored in \restore{} evenly distributed across all PEs in a way that no PE loads the same data it originally submitted.

By decreasing the size of permutation ranges (Section~\ref{sec:permutations}) we can control how many PEs can participate in sending requested data: When using smaller permutation ranges, more PEs are able to serve parts of the requested data to the requesting PEs.
Small permutation ranges, on the other hand, lead to fragmentation of the data and therefore induce many small messages.
In \Cref{fig:permutation-range-size} we show the number of bytes per permutation range on the \(x\)-axis and the running times of \submit{} and \loadone{} on the \(y\)-axis for different numbers of PEs.
We do not show results for \loadall{} because permutations even have a negative effect on performance here (\Cref{fig:id-randomization-on-or-off}).
We therefore recommend turning them off when using a recovery mechanism which loads all data stored in \restore{}.
We observe that for few bytes per permutation range, both \submit{} and \loadone{} are slower by up to an order of magnitude than the fastest configuration because of a high bottleneck number of messages.
Approaching \SI{16}{\mebi\byte} of data per permutation range, fewer PEs can participate in sending data.
Between these two extremes, there is a range of permutation range sizes which yield fast running times.
For all further experiments, we thus fix the amount of data per permutation range to \SI{256}{\kibi\byte} (\SIrange{0.65}{2.27}{\milli\second} for \loadone{} on \SIrange{48}{6144} PEs).
For \SI{16}{\mebi\byte} of data per PE, every PE receives approximately \SI{164}{\kibi\byte} in \loadone{} which results in an average of two PEs requesting the same permutation range and therefore induces a sparse communication pattern.
On the sending side, this implies that the data submitted by a single PE is distributed among \(64\) permutations ranges. With \(r=4\) redundant copies this results in up to \(64 \cdot 4 = 256\) PEs that participate in serving this part of the data.

As expected, enabling random permutations speeds up \loadone{} and slows down \loadall{}, especially for runs on many PEs (\Cref{fig:id-randomization-on-or-off}).
This is because in \loadall{}, even without permutations, every PE sends some part of the data.
By enabling permutations, the data requested by a PE is distributed among more sending PEs, resulting in a denser communication pattern.
We can tolerate an increase in running time of \submit{} as it is called only once in the case of only submitting input data. In contrast, a load is issued after every failure.

\subsection{Applications}\label{sec:applications}

\begin{figure}[!t]
    \centering
    \includegraphics{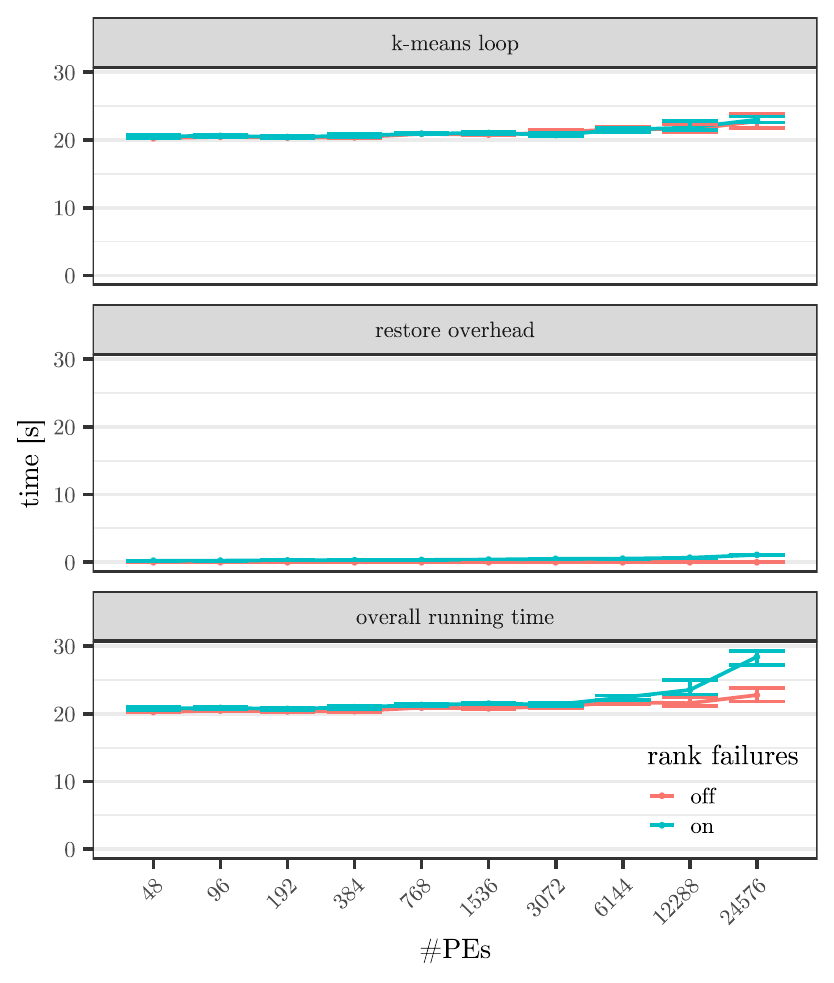}
    \caption{
        Running time of the \(k\)-means clustering algorithm with and without failures on \SI{16}{\mebi\byte} of data per PE\@.
        \emph{\(k\)-means loop:} time spent for the core clustering algorithm.
        \emph{Restore overhead:} time spent in \restore{}'s functions.
        \emph{Overall running time} also includes additional work required for attaining fault-tolerance, such as a load balancer to determine how to redistribute data and MPI functions for identifying the failed PEs.
        \label{fig:k-means}
    }
\end{figure}

\begin{figure*}[!t]
    \centering
    \subfloat[Real world datasets\label{fig:raxml-real}] {
        \includegraphics[valign=t]{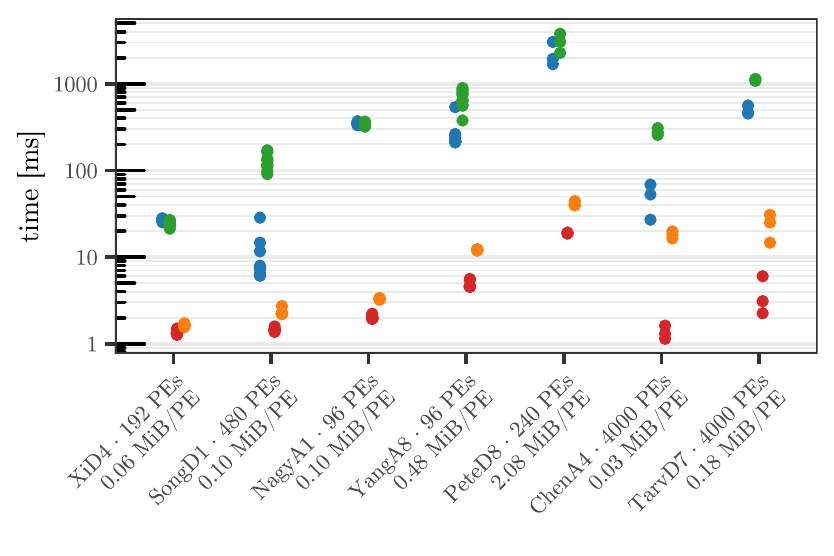}
    }
    \subfloat[\SI{19.1}{\gibi\byte} synthetic dataset\label{fig:raxml-synthetic}] {
        \includegraphics[valign=t]{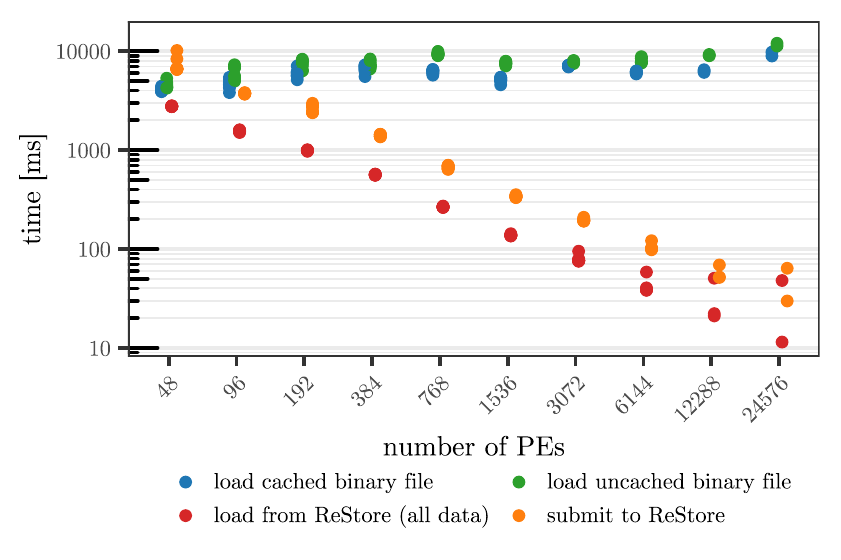}
    }
    \caption{Performance of data loading after a fault in FT-RAxML-NG.
        In subplot~(a), labels on the \(x\)-axis show the name of the data set, the number of PEs used for that data set, and the corresponding amount of input data per PE\@.\label{fig:raxml}}
\end{figure*}

To demonstrate realistic use cases of \restore{} we use it to restore lost data
in two different real-world applications.
\Cref{fig:k-means} shows running times for a small example application that computes a \(k\)-means clustering~\cite{macqueen1967}.\footnote{We ran these experiments with IntelMPI, because its \texttt{Group\_*}-functions -- which we use to determine which PEs failed -- perform better than OpenMPI's.}
Each PE holds \num{65536} points in a \(32\)-dimensional space as input with \(8\) byte double precision floating point values per dimension resulting in \SI{16}{\mebi\byte} of input data.
All PEs start with the same \(20\) random starting centers.
Iteratively, each PE assigns the nearest center to each of its local points and all PEs collaboratively calculate new centers positions using an all-reduce-operation over \(k\) elements.
If a PE fails, the remaining PEs divide the dead PE's data points evenly among them using \restore{} and continue with the calculation.
We perform \(500\) iterations of the algorithm and simulate an expected failure of \SI{1}{\percent} of all nodes distributed uniformly at random during these iterations. This is done by determining a suitable probability for each PE to fail in each iteration
of the algorithm.\footnote{Using a discrete exponential decay with \SI{1}{\percent} of failed PEs after 500 iterations.}
We find that \restore{} accounts for only \SI{1.6}{\percent} (median) of the overall running time on up to \num{24576} PEs with up to \num{262} PEs failing.
Note that the overall running time increases by more than \restore{}'s overhead for large PE counts mainly due to MPI operations used to restore a functioning communicator after a PE failure.

Next, we demonstrate \restore{}'s performance for the fault-tolerant version of the highly complex and widely used phylogenetic tree inference software RAxML-NG~\cite{Kozlov2019} -- called  FT-RAxML-NG~\cite{Huebner2021}\footnote{\url{https://github.com/lukashuebner/ft-raxml-ng/tree/restore-paper}} -- using the same empirical datasets as in~\cite{Huebner2021} (\Cref{fig:raxml-real}).
Additionally, we use a \SI{19.1}{\gibi\byte} synthetic dataset~\cite{aberer2014exabayes} for scaling experiments (\Cref{fig:raxml-synthetic}).
FT-RAxML-NG redistributes its input data among all surviving PEs.
We therefore deactivate permutation ranges for this application.
We compare \restore{}'s performance against FT-RAxML-NG's currently implemented recovery mechanism: Loading the data from the PFS using RAxML-NG's dedicated binary file format (\texttt{RBA}) which enables rapidly reading only the required subset of the input matrix.
We distinguish between the input files being uncached by the file system (in the first read) and being cached by previous reads.
Both, submitting data to \restore{} and loading data after a failure, is faster than the original method of loading the data from files -- often by more than an order of magnitude.
On the synthetic data set, for low PE counts, submitting to \restore{} is slower than reloading from a file.
However, this is negligible because an actual phylogenetic inference on this dataset requires terabytes of memory for likelihood calculations and would therefore never run on this few nodes.
We also want to emphasize that submitting to \restore{} has to be done only once in FT-RAxML-NG, while loading has to be conducted after every failure.

\subsection{Comparison with Other Approaches}\label{sec:disks}

We now evaluate \restore{}'s performance in comparison to other checkpointing approaches.
As shown in \Cref{sec:related-work}, most checkpointing libraries store checkpoints on the PFS\@.
Most on-disk and all in-memory checkpointing libraries support only substituting recovery, i.e., no shrinking recovery, yielding a comparison with \restore{} challenging.
Additionally, as detailed in \Cref{sec:competitor-reproducibility}, to the best of our knowledge, there exists no in-memory checkpointing library which is still maintained and working to compare ReStore to.

\subsubsection{Comparing to Disk-Based Approaches}
\begin{figure}[!t]
    \centering
    \includegraphics{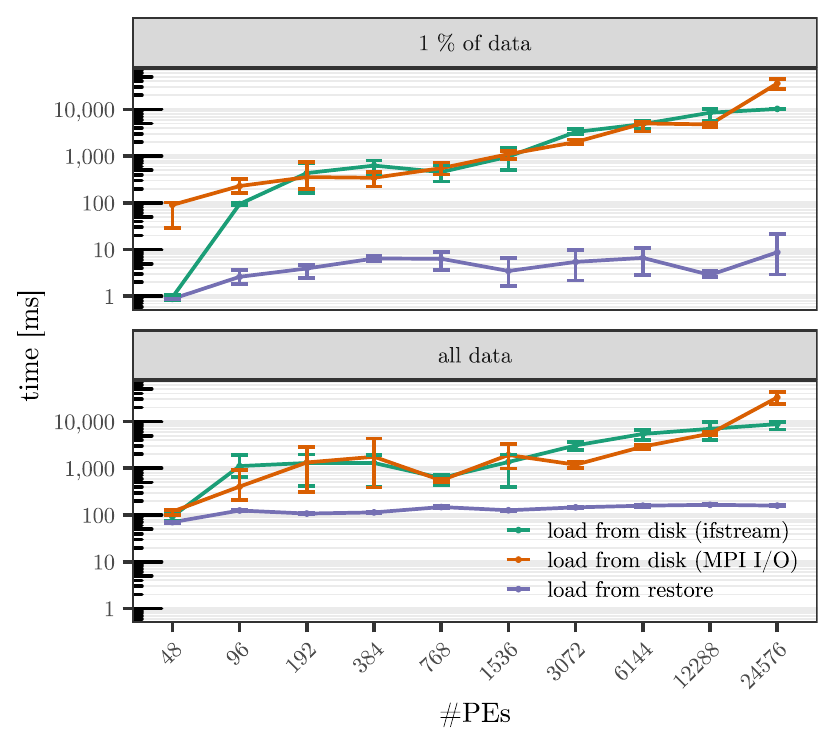}
    \caption{Loading performance of \restore{} vs.\ loading from files on the
        clusters parallel file system, representing the approach of most checkpointing libraries.}\label{fig:load-from-disk}
\end{figure}

We compare \restore{} against loading a copy of the data stored on the PFS (\Cref{fig:load-from-disk}).
We create this file such that reading is a single consecutive read and therefore as fast as possible.
We show running times for reading a separate file for each reading PE using C++'s \texttt{ifstream} and reading a single file for all PEs with \texttt{MPI\_File\_read\_at\_all} (MPI I/O in the plot).
This is a lower bound for all checkpointing libraries that have to read their data from disk.
\restore{} outperforms disk access (\texttt{ifstream}) on \num{24576} PEs by a factor of 206 (\SI{1}{\percent} of data; median) and 55 (all data) respectively.

\subsubsection{Comparing to Reported Measurements}

Gamell~\etal{}~\cite{Gamell2014} measure approximately \SI{115}{\milli\second} to write a checkpoint with \SI{14.8}{\mega\byte} per rank on \num{1000} ranks using Fenix.
Fenix implements a replication level of \(r=1\).
This means, that there exists a single copy of the data in addition to the data the ranks are actively working on.
According to our definition (\Cref{sec:probability-of-idl}), a single rank failure will cause irrecoverable data loss.
This works in practice, as long as the data which resided on the failed rank(s) does not need to be restored.
To serialize and store \SI{16}{\mebi\byte} per rank on \num{1536} ranks (32 nodes) with a replication level of \(r=1\) and using consecutive IDs, \restore{} needs \SI{126 \pm 3}{\milli\second} (\(\mu \pm \sigma\), 10 repeats).
Gamell~\etal{}~\cite{Gamell2014} expect Fenix's recovery time to be the same as its checkpointing time but do not provide experimental results for that claim.
\restore{} restores the data of a single rank to another single rank in our experiments in \SI{21 \pm 2}{\milli\second}.
\restore{} additionally offers to restore the data of a single rank scattered to all surviving ranks.
This operation requires \SI{20 \pm 5}{\milli\second} in our experiments.
In the case that one expects more than one recovery per checkpoint \restore{} offers ID permutations to speed up recovery at the cost of slower checkpoint creation (\Cref{sec:permutations}).
This would for example be the case for static input data, of which multiple ranks need different but small fractions after getting assigned new work following a rank failure.
With ID permutations enabled, saving the data to \restore{} takes \SI{215 \pm 9}{\milli\second} in our experiments.
Restoring the data takes \SI{15 \pm 3}{\milli\second} if restoring all the data to a single rank and \SI{0.9 \pm 0.2}{\milli\second} if restoring the data scattered across the surviving ranks.
As the latter evenly distributes the data across the surviving ranks, we expect it to become the more common scenario when working in a shrinking setting.

Bartsch~\etal{}~\cite{Bartsch2017} report \gpicp{} to require approximately \SI{1}{\second} to initialize, \SI{200}{\milli\second} to create a checkpoint and \SI{15}{\milli\second} to restore data from a checkpoint.

Fenix's performance was measured on a Cray XK7 system with 16 cores per node and a \SI{160}{\giga\byte\per\second} network~\cite{Cray2013}.
\gpicp{}'s performance was measured on an unnamed system with 16 cores per node and a QDR Infiniband network.
We measured \restore{}'s performance on the SuperMUC-NG, which has 48 cores per node and an OmniPath interconnection with \SI{100}{\giga\bit\per\second} (\Cref{sec:environment}).
We choose our experiments such that data is always copied between different nodes and never between two processes running on the same node.
Thus, all 48 processes on a single node have to share the same interconnect. 
Considering that we evaluate \restore{} on a slower network than Fenix, we expect an even more favourable comparison when having access to a similar HPC system.

Lu~\cite{Lu2005} reports checkpoint creation times of \SIrange{8}{20}{\second} for \SIrange{157}{182}{\mega\byte} on \num{448} ranks.
They report restoration times of \SIrange{20}{48}{\second}.
Thus, assuming linear scaling, we expect checkpoint creation times of approximately \SI{1}{\second} and restoration times of approximately \SI{2}{\second} for \SI{16}{\mebi\byte} of data.
Lu's algorithm is thus an order of magnitude slower than \restore{} and Fenix.
We assume this is due to the fact, that Lu's algorithm uses erasure codes (\Cref{sec:memory-usage}).

To summarize, \restore{} can be configured to create and restore from checkpoints in the same manner and approximately the same time as existing checkpointing solutions.
\restore{} additionally has functionality to (a) increase the replication level (b) restore the data in a scattered manner to multiple ranks instead of to one rank and (c) enable ID permutations to decrease time to restore the data by an order of magnitude while doubling the time taken to create a checkpoint.
The latter option is for example useful when creating a replicated storage for the input data of a program, which has to be partially reload after a failure.

\section{Conclusion and Future Work}\label{sec:conclusion}
We show that by using a suitable data distribution strategy, recovery of lost data after a failure is possible in tens to hundreds of milliseconds, depending on the amount of data loaded.
We achieve this by using a distribution scheme for redundant copies that ensures a low probability of data loss and a rapid recovery of the data.
We also provide the -- to the best of our knowledge -- first in-memory checkpointing library which supports shrinking recovery, that is \restore{} is able to restore the data of the failed PEs scattered to multiple or all surviving ranks instead of to a single respawned or spare PE.
This alleviates the need for the application to allocate spare nodes which participate in the computation only in case of a node failure, thus increasing computation efficiency.
We supply an analysis of the probability of irrecovable data loss and propose a data distribution to easily restore lost replicas after a failure.
Experimental and theoretical evaluation of the proposed data redistribution after a node failure constitutes part of future work.
This further decreases the probability to lose all copies of any data.
With our C++ library, we were able to improve recovery performance of FT-RAxML-NG~\cite{Kozlov2019, Huebner2021} by up to two orders of magnitude.
By using the proposal implementation of the fault-tolerance mechanisms included in the new MPI 4.0 standard, our library can be used by applications on HPC systems once the new standard is implemented.
We also plan on evaluating \restore{} for checkpointing of dynamic program state and extend its API for different data formats (e.g\@. 2D data).

\FloatBarrier{}
\section*{Acknowledgments}
The authors gratefully acknowledge the Gauss Centre for Supercomputing e.V\@. (\url{www.gauss-centre.eu}) for funding this project by providing computing time on the GCS Supercomputer SuperMUC-NG at Leibniz Supercomputing Centre (\url{www.lrz.de}).
Part of this work was funded by the Klaus Tschira foundation.
This work was supported by a grant from the Ministry of Science, Research and the Arts of Baden-Württemberg (Az: 33-7533.-9-10/20/2) to Peter Sanders and Alexandros Stamatakis. 
This project has received funding from the European Research Council (ERC)\footnote{\includegraphics[scale=0.05]{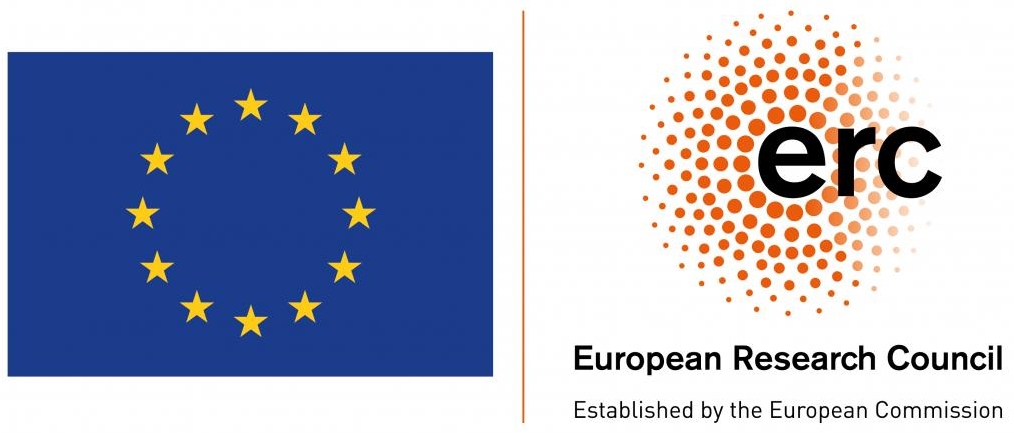}} under the European Union's Horizon 2020 research and innovation program (grant agreement No. 882500).
We thank an anonymous reviewer for pointing out that the simple approximation of the irrecoverable data loss given in \Cref{sec:probability-of-idl} is very accurate for small $f$.

\appendix[Choosing a Random Permutation in \Cref{sec:restore_redundancy}]
\label{sec:choosing-the-hash-function}

\subsubsection{Data Distribution A}

\begin{figure}[!t]
    \centering
    \includegraphics{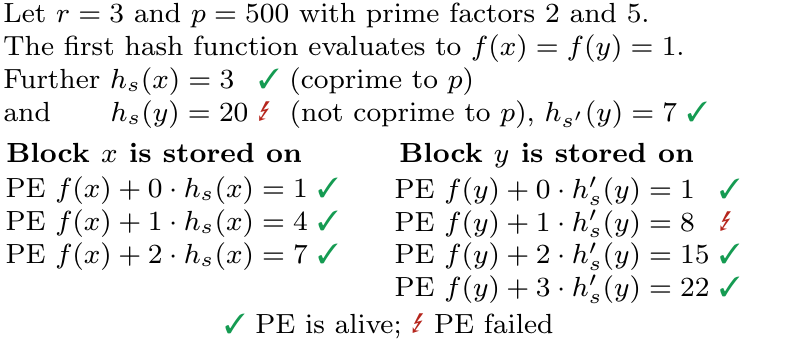}
    \caption{A modified data distribution enabling easy recovery of replicas (Data Distribution A). \(r\) denotes the number of replicas, \(p\) the number of nodes, \(f\) and \(h\) are hash functions.\label{fig:restore-redundancy-a}}
\end{figure}

Let \(f\) and \(h_s\) be fast-to-compute hash functions that avoid collisions, where \(s\) is a seed used to parametrize \(h_s\).
We compute the nodes where a block \(x\) is located by evaluating \(\rho_x(k) = (f(x) + k \cdot h_s(x)) \mod p\) for \(k = 0, 1, \ldots \) until we find \(r\) nodes that are still alive (\Cref{fig:restore-redundancy-a}).
This scheme is analogous to collision resolution in hash tables using open addressing and double hashing.
The probing sequences \(\rho_x\) and \(\rho_y\) are likely to be different for two blocks \(x\) and \(y\), even if \(\rho_x(0) = \rho_y(0)\).

If \(L(x, k) = L(x, j)\) for \(k < j < p\), the entire sequence will repeat from that point on, yielding recovery of lost replicas  impossible after more than \(j\) failures.
We can avoid this by requiring the seeded hash function \(h_s\) to only yield integers that are coprime to \(p\).
We archive this by trying different seeds \(s\) for \(h_s\) until \(h_s(x)\) is coprime to \(p\).
For this purpose, we draw a sequence of seeds at program startup.
The probability of two random integers being coprime is \(6/\pi^2\)~\cite{Hardy1960}.
For a random\footnote{Though realistically, we cannot expect $p$ to be random.} \(p\), the expected number of differently seeded hash function we have to evaluate is thus
\[
    1 + \sum_{n=1}^\infty \left(1 - \frac{6}{\pi^2} \right)^n = \frac{7}{6} \left(\pi^2 - 6 \right) \approx 1.65
\]
To check if \(h_s(x)\) is coprime to \(p\), we divide \(h_s(x)\) by every prime factor of \(p\), which we factorize once during program startup.
The Erdős-Kac theorem~\cite{Erdos1940} states that the number of distinct prime factors \(m\) of a random number below \(\hat{p}\) approximately follows the normal distribution with a mean and variance of \(\log \log \hat{p} \).
For example, a node count of \(p < 10^9\), has on average \(m = 3 \pm{} 1.7\) prime factors.
For node counts \(p < 10^9 \) (\(m=3\)) we therefore expect \(< m \cdot 1.65 = 5\) divisions to check for coprimality each time we need to compute on which nodes to store a block.

\subsubsection{Data Distribution B}

Alternatively, we can deploy a classical seeded pseudorandom permutation \(\rho_s\) on \([0, p-1]\) to generate an independent probing sequence for each element \(x\).
Let \(f\) be a collision  avoiding hash function.
We then use \(f(x)\) as the seed for \(\rho\).
As \(\rho\) we could for example use a Feistel Network-based permutation with \(f(x)\) as the seed and cycle walking for values of \(p\) which are not a power of two.

\bibliographystyle{IEEEtran}
\bibliography{references}
\end{document}